# Evaluating Heuristics for Iterative Impact Analysis

Yibin Wang
Department of Computer Science
Wayne State University
Detroit, USA
yibin.wang@wayne.edu

Maksym Petrenko
Earth System Science Interdisciplinary Center
NASA/GSFC
Greenbelt, USA
maksym.petrenko@nasa.gov

Václav Rajlich
Department of Computer Science
Wayne State University
Detroit, USA
rajlich@wayne.edu

*Abstract*—Iterative impact analysis (IIA) is a process that allows developers to estimate the impacted units of a software change. Starting from a single impacted unit, the developers inspect its interacting units via program dependencies to identify the ones that are also impacted, and this process continues iteratively. Experience has shown that developers often miss impacted units and inspect many irrelevant units.

In this work, we study propagation heuristics that guide developers to find the actual impacted units and termination heuristics that help to decide whether the estimated impact is complete. The roles of these two kinds of heuristics are complementary and affect both the precision and recall when used during IIA. We investigated several propagation heuristics adapted from previously published papers and combined them with a practical termination heuristic. We developed a reenactment process that simulates the actions of developers who use those heuristics during IIA, and we assessed their performance. The software changes for our reenactment were mined from the repositories of open source projects. We found that IIA provides better recall than the other known impact analysis techniques. However the IIA with the propagation heuristics that we investigated does not supersede IIA combined with a random inspection, and hence these heuristics do not help the IIA.

## I. INTRODUCTION

During software evolution, developers continuously make software changes. Impact analysis (IA) is a design phase for a software change where developers estimate which units should be modified [1].

*Iterative impact analysis (IIA)* is a process that allows developers to detect impacted units step by step following program dependencies of a software system. It starts with an initial impacted unit that is scheduled to be modified; this unit can be identified during a preceding phase named concept location [2]. The developers inspect other units that interact with the initial impacted unit and determine whether these units are impacted by the change also. This process continues iteratively and has been called ripple effect [3].

All units inspected by the developers during IIA constitute the *visited set (VS)*, the units that developer predict to be modified form the *estimated impact set (EIS)*, and the *actual impact set (AIS)* consists of all units that are modified in the real implementation. The estimated impact set may contain units that are not modified finally; such units are false positives of IA. It is also possible that the estimated impact set misses some units of the actual impact set, which are false negatives of IA. Numerous IIA techniques have been investigated in the past [4-7].

Some researchers proposed IA techniques that predict the EIS in a single algorithmic step [8-11]. We call this process *all-at-once impact analysis (AIA)*. Compared to AIA, developers are "in the loop" during IIA and are expected to correct imperfections that an AIA algorithm accumulates.

In this paper, we investigate IIA and two kinds of heuristics that may assist it. One type is *propagation heuristics* that guide the developers towards the units that are likely to be impacted by the change. They play a role in the situation where there are many interacting units for the developer to inspect. They guide developers towards the units that are most likely to change. The other type of heuristics is *termination heuristics* that indicate that the EIS is complete. The roles of these two kinds of heuristics are complementary and affect both the precision and the recall for the assisted IIA technique.

In our study, we evaluate the performance (i.e. the precision and the recall) of IIA techniques that are based on a static dependency graph that is enriched by selected propagation and termination heuristics. Some of the propagation heuristics have been investigated within the context of AIA techniques [8, 10, 12] and we adapted them for IIA. In order to compare the effectiveness of these heuristics, we developed an empirical approach called *reenactment* that simulates the actions of developers who are guided by heuristics during IIA. This reenactment is applied to the past changes of open source projects mined from software repositories.

The rest of the paper is organized in the following way: Section II summarizes the previous work. Section III describes the static dependency graph and the heuristics of this study. In section IV, we present the design of our reenactment that we use to compute the performance of the heuristics. Section V contains the results, our findings and threats to validity. In section VI, we discuss conclusions and future work.

## II. PREVIOUS WORK

Arnold and Bohner presented an early classification of impact analysis techniques [13]. Lehnert developed a taxonomy that includes the scope of analysis, the granularity of units, utilized techniques, style of analysis, tool support, supported languages, scalability, and experimental results [14]. Li et al. focused on code-based IA techniques and also categorize them by several properties [15].

Researchers have integrated different IA techniques to enhance the overall performance over the past decade.

Malik and Hassan [16] combined machine learning with several IA techniques to predict the EIS for C programs. The best result was 78% recall and 64% precision. Several approaches were proposed to divide the IA into several subtasks and applied different IA technique in each subtask [17-19]. Among them, Zanjani et al. [17] reached 32% recall and 12% precision for IA at the granularity of files, when the size of VS was 20. Borg et al. [19] aimed at predicting the impact on non-code artifacts, and they got 60% recall and 7% precision when the size of VS was 40. Some researchers investigated the union and intersection of the EIS found by different IA techniques to improve the performance [9, 10], but the recall was still low. Sun et al. [12] compared three tools based on different static IA techniques and studied the union and intersection of the results from each tool. They achieved 61% recall and 38% precision at the granularity of classes. Musco et al. [20, 21] proposed two propagation heuristics based on the execution of test cases to predict how a change in the production code impacts harness code. This resulted in 79% recall and 69% precision at the granularity of methods. All these publications studied AIA techniques.

In the literature, IA techniques based on program slicing are believed to have high recall, though they are very costly in practice [23, 24]. Several approaches were proposed to approximate program slicing at the granularity of methods with lower cost [25, 26]. However, Toth et al. [6] found that at the granularity of classes, program slicing has very low recall (i.e. 11.65%) and does not meet the needs of developers.

The IIA process has been described in [2]. It uses the marks in Table I. The marks indicate the status of the program units during IIA. They include the 'Propagating' mark that is used for units that are not changed, but still propagate a change to their interacting units [5, 7]. Both 'Propagating' and 'Unchanged' units are inspected by developers, but they are treated as the false positives that increase the workload of the developers and lower the precision of impact analysis. Iterative IA theoretically finishes when there is no unit scheduled to inspect (marked as 'Next'). However, in practice, the set of 'Next' units is often large, so developers terminate the process as soon as they conclude that EIS has been identified [27]. In this situation, the order in which the developers inspect the 'Next' units should impact the number of false positives of IIA, and hence the prioritization of those units becomes an issue.

The Class and Member Dependency Graph (CMDG) and its implementation, a tool JRipples [28], use these marks to support IIA in Java projects [6, 7, 29]. Past results showed that IIA supported by JRipples can reach 100% recall [5, 7], though the precision was low. For a system with 500 classes, JRipples usually finishes building the CMDG in a minute. In this context, we evaluate selected heuristics at the granularity of classes.

In the work of Petrenko [30], several propagation heuristics were investigated. That work compared the average precision of IIA tasks when the recall is 100%. Instead of this assumption, we introduce termination heuristics. We investigate both the precision and recall for a set of propagation heuristics combined with the termination heuristic. This makes the reenactment more realistic, compared to that in [30].

Abi-Antoun et al. [7] proposed another IIA technique based on a hierarchical object graph that provides both the program dependencies for developers to follow as well as a propagation heuristic to rank these dependencies. However, generating this hierarchical object graph requires non-trivial manual effort [22].

TABLE I. MARKS IN IIA

| Mark | Meaning |
| --- | --- |
| Blank | Unknown status of the unit; the unit was never inspected and is not currently scheduled for inspection. |
| Impacted | The developers inspected the unit and found that it was impacted by the change, i.e., the unit belonged to the EIS. All 'Blank' neighbors of this unit must be scheduled for inspection, i.e., they are marked 'Next'. |
| Unchanged | The developers inspected the unit and found that it was not impacted by the change. This unit does not propagate the change to any of its neighbors. |
| Next | The unit is scheduled for inspection by the developers. |
| Propagating | The developers inspected the unit and found that it was not impacted by the change, but the neighbors of this unit may still change. All 'Blank' neighbors of this unit must be marked 'Next'. |

## III. DEPENDENCY GRAPH AND HEURISTICS

The class dependency graph is defined in the following way.

**Definition 1.** *Let P be a program and let G = (V, E) be a directed graph where V is the set of all classes in P and E is the set of directed edges. An edge (x, y) ∈ E if and only if the class y or any member of y is referred (e.g., called, inherited, extended, instantiated, etc.) within the definition of the class x. Then, G is the Class Dependency Graph (CDG) of the program P.*

For impact analysis, we use the symmetric closure for the edges in CDG because the change can propagate in both directions through an edge (i.e. dependency) and we add supports for propagation heuristics.

**Definition 2.** *Let G = (V, E) be a CDG, then $G_H$ = (V, E', $W_H$) is a Weighted Class Propagation Graph (WCPG) with a set of classes V, a set of edges E' where E' is a symmetric closure of E, and a set of weights $W_H$ where each weight $w_H(x, y)$ is produced by a propagation heuristic H for an edge (x, y) ∈ E'. Note $w_H(x, y)$ could be different to $w_H(y, x)$. We say y is a neighbor of x if there is an edge(x, y) ∈ E'.*

During IIA, if a class *x* is marked as 'Impacted' or 'Propagating', $w_H(x, y)$ will be determined by a specific propagation heuristic to rank its neighbor *y*. We assume that a higher ranked neighbor is more likely to change or propagate the change.

*A. Propagation Heuristics*

We selected several representative heuristics for our study:

*1) Dependency Based Heuristic (DBH)*

An extensive survey of heuristics based on static dependencies was conducted in [31], and the PIM heuristic (i.e. the number of method invocations between classes, taking into account the polymorphism) had particularly good performance among them. PIM was also used in more recent IA studies [12, 32]. Thus, we derive our DBH from this heuristic. If call(x,y) denotes the number of times the class x calls any method of the class y, including polymorphic calls, then DBH(x,y) = call(x,y) + call(y,x). The value of DBH(x, y) is always a natural number.

*2) Class to Class similarity by Information Retrieval (CCIR)*

Latent Semantic Indexing is an information retrieval method that constructs a vector space model (VSM) for texts where each text is represented by a vector. Before constructing VSM, the source code of classes is pre-processed to identify meaningful words; this may include splitting composite identifiers, removing language-specific stop-words, and so forth. A non-negative cosine value of the angle between the corresponding vectors of two texts in VSM indicates their similarity and serves as the conceptual coupling [10, 32]. Our CCIR(x, y) is the non-negative cosine value of the angle between the vectors representing classes x and y.

*3) Change Request to Class similarity by Information Retrieval (RCIR)*

A change request is a text that describes the required modification to the program. The heuristic in [33] compares the text of a change request to the text in the source code of classes, and is based on the assumption that terms appearing in the text of the change request also appear in the source code of the impacted classes.

Let IR(r, x) be the non-negative cosine value of the angle between the vectors representing a class x and a change request r. We adjust it as a propagation heuristic in the following way: For every edge from a class x to a class y in $G_H$, RCIR(x, y) = IR(r, y).

*4) Evolutionary Coupling between Classes by Mining Software Repositories (Hist1 and Hist2)*

Mining Software Repositories (MSR) methods can uncover unique relations for change propagation among program units [34], which are not detectable by program analysis. Such relations are called evolutionary couplings [9, 10]. Association rules are specific kinds of evolutionary couplings between a resource unit m and a destination unit n determined by a set of commits in a training set. Each rule comes with a *support value*, which indicates how frequently both m and n appear together in a single commit in the training set, denoted by $Asso_S(m, n)$. There is also a *confidence value*, which indicates how often the commits containing m also contain n, denoted by $Asso_C(m, n)$. $Asso_S(m, n)$ is always symmetric but $Asso_C(m, n)$ can be different from $Asso_C(n, m)$. Zimmermann et al. explored an AIA technique based on association rules [35]. That is, for the given initial impacted unit u, all association rules using u as the resource unit with a non-zero support value are collected. Then the destination units of those rules construct the EIS and are ranked by the corresponding confidence value. This AIA technique was combined with other IA techniques in more recent studies [9, 10, 12].

In our study, we extract change history from a selected period of commits in the repositories as the training set. Then we build association rules among all classes. Next, for an edge from a class x to a class y in $G_H$, we investigate two propagation heuristics Hist1(x, y) = $Asso_C(x, y)$ and Hist2(x, y) = $Asso_S(x, y)$.

*5) Random Propagation Weight between Classes (RND)*

We add RND, i.e. random weights ranging from 0 to 1, for all edges in $G_H$. RND is used as the baseline for assessing the performance of all propagation heuristics in order to show whether they are better than a completely random inspection. We generate RND weights only once for each subject system in our experiment and reuse those weights to reenact all cases of that system.

*B. Termination Heuristics*

Our termination heuristic, denoted by TopN, is based on the idea that a developer would inspect no more than N neighbors of every 'Impacted' or 'Propagating' class. Those are the neighbors that are ranked highest by the specific propagation heuristic that we are exploring. Similar heuristic, cut point, was used in many research papers such as [8, 10, 12]. Without termination heuristics, programmers have to inspect all neighbors iteratively from the initial impact set until all reachable classes have been inspected.

TopN is defined by the following way:

**Definition 3.** *Let x be a class in $G_H$, then neighbors(x) = { y | (x, y) ∈ E' } and weights(x) = {$w_H(x, y)$ | y ∈ neighbors(x) }.*

**Definition 4.** *Given a natural number N and a class x in $G_H$, let y∈neighbors(x) and $w_H(x, y)$ is the N'th largest weight in weights(x), then TopN(x) is a set of reachable neighbors for x where TopN(x) = {v | v∈neighbors(x) such that $w_H(x, v) ≥ w_H(x, y)$ }.*

In order to adjust this heuristic to the subject systems of different size, we decided to use the percentage of the total number of classes in the subject system to determine the actual value N of the heuristic. We selected different percentages in this study, which are 0.5%, 1%, 2%, 3%, 4% and 5%.

## IV. DESIGN OF THE CASE STUDY

The empirical method used in our study is reenactment. An overview of the reenactment is shown in Figure 1.

Each change request that we select for this study involves a single feature only. To achieve that, we investigate closed tickets on SourceForge to extract the description of each change request, the change ID, and the date when the change request was resolved. Then, we find the corresponding revision in the SVN repository. Next, we parse the change details of that revision to extract the AIS and ensure it fulfills a single change request criterion. This part of work is done manually.

After mining the change request and its AIS, we check out the source code before the revision and use JRipples and a propagation heuristic to generate the WCPG for that code.

The repositories do not provide information on the 'Propagating' and 'Unchanged' classes that were inspected during the IA by the programmers. We reconstruct the sets of these classes by an algorithm that simulates actions of the programmers during the IIA.

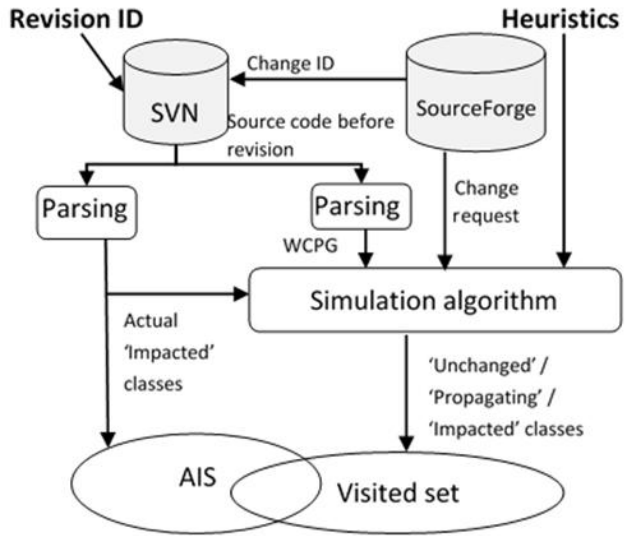

Figure 1. Overview of the IIA Reenactment

### A. Simulation Algorithm

The simulation algorithm used in our reenactment consists of the following steps:

*1) Select the initial impacted class (IIC)*

From the information in the repositories and SourceForge, the actual starting point of the change is not available. Thus, for a software change that has multiple classes in its AIS, we repeat the simulation by selecting each class of the AIS as the IIC.

*2) Build a subgraph of WCPG based on the IIC and the termination heuristic*

In the WCPG weighted by a selected propagation heuristic, we use the IIC as the root and construct a subgraph of WCPG such that it contains all classes and edges reachable from IIC in WCPG after applying a specific TopN:

**Definition 5.** *Let $G_H$ = (V, E', $W_H$) be a WCPG, c∈V is the IIC and N is a natural number, then $G_c$ = ($V_c$, $E_c$, $W_c$) is a weighted subgraph based on c with a set of reachable classes $V_c$ for c where $V_c$ = { x | x=c or there exists y∈$V_c$ such that x∈TopN(y)}. The set of reachable edges $E_c$ is defined as $E_c$ = { (x, y) | there exists x∈$V_c$ and y∈TopN(x) such that (x, y)∈E' }, and for the edge (x, y)∈ $E_c$ $w_c(x, y) = w_H(x, y)$.*

As an example, a $G_H$ weighted by propagation heuristic is shown in the left part of Figure 2 where the members of AIS have the black filling. The initial impacted class is indicated as 'IIC'. Suppose N=2, then the right part of Figure 2 shows the corresponding $G_c$ that contains all the reachable nodes and edges of the IIC after applying the termination heuristic Top2. It can be seen that some classes of the AIS are no longer reachable from the IIC, due to termination heuristics.

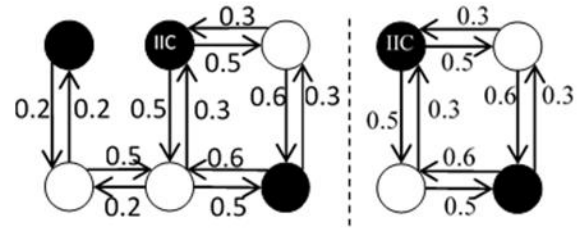

Figure 2. Finding reachable nodes and edges in WCPG based on the IIC and the termination heuristic Top2

*3) Identification of Propagating Classes*

After constructing $G_c$, the intersection between AIS and $V_c$ represents the *reachable part of AIS* for the given propagation and termination heuristics. Then the reenactment algorithm simulates the inspections that the original developer made.

If the reachable part of AIS is disconnected, the developer must have visited 'Propagating' classes during IIA. The reenactment assumes that the developer visited the minimal number of 'Propagating' classes. For the simulation algorithm, this is equivalent to resolve the graph-theoretical directed Steiner tree problem in a weighted directed graph [36] where all edges share identical weight.

**Definition 6.** Let $G_c = (V_c, E_c, W_c)$ be a weighted subgraph rooted on $c \in V_c$, then $G'_c = (V_c, E_c)$ is a converted graph from $G_c$ with the same set of classes $V_c$ and the same set of edges $E_c$ as $G_c$ and identical weight on each edge.

Then, the graph-theoretical Steiner tree problem is formulated in the following way: given $G'_c$ and $M = V_c \cap AIS$, find the sub-tree $T$ of $G'_c$ where the root $c$ has a path to every node in $M$ and the sum of the weights on the paths is the minimum. Note that $T$ may include several interconnecting nodes that are not in $M$; these nodes are known as the Steiner nodes.

As an example, a $G'_c$ with all weights equal to 1 is shown in the left part of Figure 3 where the nodes in the reachable AIS have the black filling. The root vertex is indicated with 'IIC'. Then the directed Steiner tree is depicted in the right part of Figure 3, and the Steiner nodes are indicated by the letter 'P'.

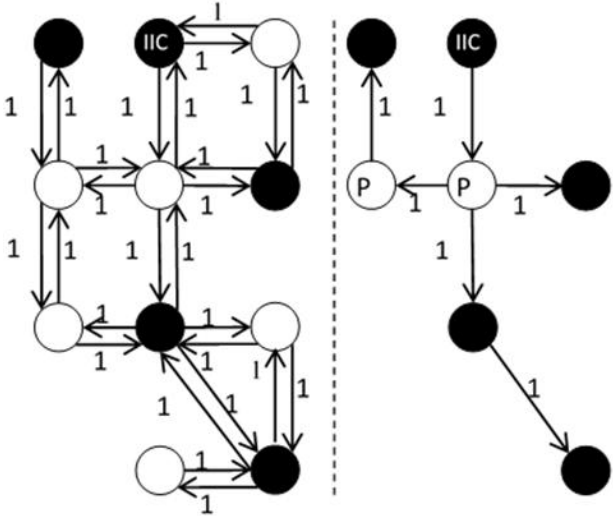

Figure 3. Finding Steiner nodes example. Left part is the directed graph, and the right part is the directed Steienr tree accordingly.

Since the problem of resolving directed Steiner tree is NP hard, we use an approximation solution [37] taking $G'_c$ and the reachable AIS as the input.

*4) Reenactment of the visited set*

When using an IIA technique, the visited set contains all inspected units, i.e. 'Impacted', 'Propagating' and 'Unchanged' units. The simulation algorithm uses $G'_c$ to compute the visited set. For every 'Impacted' or 'Propagating' class (i.e. every class in the directed Steiner tree), it marks all neighbors in $G'_c$ as "Unchanged." The assumption is that the developer inspected all these units because of the guidance by both propagation and termination heuristics.

*B. Subject Systems*

Our study is conducted on three different open source Java projects, as summarized in Table II.

To get association rules used by heuristics Hist1 and Hist2, we select all commits in SVN repository during a certain date interval, which is shown by the columns Date Interval and #Commits of Table II.

All change requests of our study were originally done after the selected date interval for association rules. We have extracted 15 change requests that have been applied on JEdit (https://sourceforge.net/projects/jedit/), 10 requests applied for JHotdraw (https://sourceforge.net/projects/jhotdraw/), and 11 requests applied for QuickFIX/J (https://sourceforge.net/projects/quickfixj/), respectively. The AIS of each change involves at least 2 classes and up to 7 classes.

*C. Measures*

For each change request $cr$ with a specific initial impacted class $c$, we collect its precision P(cr, c) and recall R(cr, c):

$$P(cr, c) = \frac{|VS \cap AIS - c|}{|VS - c|} \quad (1)$$

$$R(cr, c) = \frac{|VS \cap AIS - c|}{|AIS - c|} \quad (2)$$

Next, the precision P(cr) and recall R(cr) for each change request is the average of precisions and recalls calculated for each possible initial impacted class:

$$P(cr) = \sum_{c \in AIS} \frac{P(cr, c)}{|AIS|} \quad (3)$$

$$R(cr) = \sum_{c \in AIS} \frac{R(cr, c)}{|AIS|} \quad (4)$$

For each subject system, let CR denote the set of its change requests, then we measure the average precision $P_{avg}$ and average recall $R_{avg}$:

$$P_{avg} = \sum_{cr \in CR} \frac{P(cr)}{|CR|} \quad (5)$$

$$R_{avg} = \sum_{cr \in CR} \frac{R(cr)}{|CR|} \quad (6)$$

The results of formula (3) to (6) are discussed in the next section.

## V. RESULTS OF THE STUDY

Table III and Table IV show $P_{avg}$ and $R_{avg}$ of investigated heuristics of each subject system, along with the corresponding standard deviations according to involved P(cr) and R(cr). The column Act of these tables lists the actual N value corresponding to a specific percentage given in the column % for a system. At the end of each table, we also compute the overall average performance of all change requests from these three systems after classifying the results by the same percentage value of TopN.

Figure 4 to Figure 7 depict how $P_{avg}$ and $R_{avg}$ changes with the increase of TopN for JEdit, JHotdraw, QuickFIX/J and overall, respectively.

In addition, Figure 8 to Figure 10 show how the median of P(cr) and R(cr) changes with the increase of TopN for JEdit, JHotdraw and QuickFIX/J, respectively.

TABLE II. SUBJECT SYSTEMS

| System | Version | LOC | Classes | History for Association Rules | | # Requests |
|---|---|---|---|---|---|---|
| | | | | Date Interval | # Commits | |
| JEdit | 4.3 | 109k | 531 | [2004-12-31, 2009-12-22] | 2051 | 15 |
| JHotdraw | 7 | 83k | 568 | [2006-11-1, 2010-8-1] | 411 | 10 |
| QuickFIX/J | 1.5.3 | 30k | 281 | [2005-2-28, 2011-11-2] | 1187 | 11 |

TABLE III. AVERAGE RECALL AND ITS STARNDARD DEVIATION OF INVESTIGATED HEURISTICS

| TopN | | RND | DBH | Hist1 | CCIR | Hist2 | RCIR | RND | DBH | Hist1 | CCIR | Hist2 | RCIR |
|---|---|---|---|---|---|---|---|---|---|---|---|---|---|
| % | Act | \multicolumn{6}{c}{JEdit - Recall (%)} | | | | | | | |
| 0.5 | 3 | 87.5 | 40 | 60.8 | 40.8 | 60.8 | 35.8 | 30 | 43 | 37 | 38 | 37 | 39 |
| 1 | 6 | 97.5 | 92.5 | 87.5 | 82.5 | 87.5 | 87.5 | 13 | 28 | 36 | 30 | 36 | 28 |
| 2 | 11 | 100 | 100 | 95 | 95 | 95 | 100 | 0 | 0 | 26 | 26 | 26 | 0 |
| 3 | 16 | 100 | 100 | 95 | 100 | 95 | 100 | 0 | 0 | 26 | 0 | 26 | 0 |
| 4 | 22 | 100 | 100 | 100 | 100 | 100 | 100 | 0 | 0 | 0 | 0 | 0 | 0 |
| 5 | 27 | 100 | 100 | 100 | 100 | 100 | 100 | 0 | 0 | 0 | 0 | 0 | 0 |
| % | Act | \multicolumn{6}{c}{JHotdraw - Recall (%)} | | | | | | | |
| 0.5 | 3 | 75 | 70.5 | 31.8 | 59.1 | 31.8 | 45.6 | 24 | 20 | 33 | 25 | 33 | 27 |
| 1 | 6 | 90.9 | 93.2 | 84.1 | 95.5 | 84.1 | 66.3 | 14 | 13 | 19 | 7 | 19 | 25 |
| 2 | 11 | 97.7 | 95.5 | 97.7 | 97.7 | 97.7 | 95.5 | 5 | 11 | 5 | 5 | 5 | 11 |
| 3 | 17 | 97.7 | 97.7 | 97.7 | 97.7 | 97.7 | 97.7 | 5 | 5 | 5 | 5 | 5 | 5 |
| 4 | 23 | 97.7 | 97.7 | 97.7 | 97.7 | 97.7 | 97.7 | 5 | 5 | 5 | 5 | 5 | 5 |
| 5 | 28 | 97.7 | 97.7 | 97.7 | 97.7 | 97.7 | 97.7 | 5 | 5 | 5 | 5 | 5 | 5 |
| % | Act | \multicolumn{6}{c}{QuickFIX/J - Recall (%)} | | | | | | | |
| 0.5 | 2 | 68.1 | 45.8 | 36 | 68.1 | 36 | 37.5 | 36 | 38 | 27 | 36 | 27 | 35 |
| 1 | 3 | 82.5 | 72.5 | 71.3 | 82.5 | 71.3 | 58.1 | 29 | 36 | 36 | 29 | 36 | 36 |
| 2 | 6 | 100 | 95 | 90 | 97.5 | 90 | 90 | 0 | 11 | 18 | 10 | 18 | 31 |
| 3 | 9 | 100 | 95 | 97.5 | 100 | 97.5 | 100 | 0 | 11 | 6 | 0 | 6 | 0 |
| 4 | 12 | 100 | 100 | 100 | 100 | 100 | 100 | 0 | 0 | 0 | 0 | 0 | 0 |
| 5 | 15 | 100 | 100 | 100 | 100 | 100 | 100 | 0 | 0 | 0 | 0 | 0 | 0 |
| % | Act | \multicolumn{6}{c}{Overall Recall (%)} | | | | | | | |
| 0.5 | - | 78.1 | 50.2 | 45.2 | 54.2 | 45.2 | 39.0 | 31 | 36 | 33 | 34 | 33 | 35 |
| 1 | - | 91.1 | 86.6 | 81.6 | 86.1 | 81.6 | 72.6 | 20 | 28 | 32 | 25 | 32 | 30 |
| 2 | - | 99.4 | 97.2 | 94.2 | 96.5 | 94.2 | 95.7 | 3 | 8 | 20 | 18 | 20 | 18 |
| 3 | - | 99.4 | 97.8 | 96.5 | 99.4 | 96.5 | 99.4 | 3 | 7 | 17 | 3 | 17 | 3 |
| 4 | - | 99.4 | 99.4 | 99.4 | 99.4 | 99.4 | 99.4 | 3 | 3 | 3 | 3 | 3 | 3 |
| 5 | - | 99.4 | 99.4 | 99.4 | 99.4 | 99.4 | 99.4 | 3 | 3 | 3 | 3 | 3 | 3 |

*A. Discussion of results*

According to Table III, IIA combined with RND provides a better recall than many IA techniques that were presented in the literature. However, when TopN is low, RND also reaches a better recall compared to the propagation heuristics that we investigated in our study. In addition, propagation heuristics based on information retrieval or evolutionary couplings lead to low recall for low TopN, especially when it is Top0.5%.

It is worth noting that in JHotdraw, the highest recall on average stops at 97.7% for any investigated propagation heuristic. This is caused by a specific change request related to Revision ID 783: "It should distinguish between large icon and small icon. This way, an Action can use different icons for buttons and menu items."

This change affects 7 classes. One class of the AIS, CrossPlatformApplication, is only interacting with another class named ResourceBundleUtil. Unfortunately, in the neighborhood of ResourceBundleUtil by every propagation heuristic including RND, CrossPlatformApplication is ranked lower than Top5%, which leads to a situation that our reenactment is not able to reach 100% recall unless using CrossPlatformApplication as the IIC.

In JEdit and QuickFIX/J, RCIR always get the better precision compared to any other propagation heuristic for the same TopN, as shown in Table IV.

Similar performance of RCIR continues in JHotdraw, However, DBH provides good precision, which is lower only than RCIR, while maintaining far better recall.

Though the confidence value and support value of an association rule are different, they rank the neighborhood of an 'Impacted' or 'Propagating' class in the same order. As a result, the simulation using Hist1 has the exactly same performance as that using Hist2 in all the cases.

Figures 4 to 7 show that for each propagation heuristic, both precision and recall become stable after TopN reaches 2% for all subject systems. This may imply that Top2% is a sufficient termination heuristic and there is no need to investigate termination heuristics that require an inspection of the larger number of neighbors.

Also note that when the value of TopN is low, reenactment requires a lot of 'Propagating' classes in order to achieve the best recall. This is why for some propagation heuristics, the precision is improved when TopN increases at the beginning.

According to Figure 8 to Figure 10, once the median of recall reached 100%, any investigated propagation heuristic leads to slightly better median precision compared to random inspection.

TABLE IV. AVERAGE PRECISION AND ITS STARNDARD DEVIATION OF INVESTIGATED HEURISTICS

| TopN | | RND | DBH | Hist1 | CCIR | Hist2 | RCIR | RND | DBH | Hist1 | CCIR | Hist2 | RCIR |
|---|---|---|---|---|---|---|---|---|---|---|---|---|---|
| % | Act | JEdit – Precision (%) | | | | | | Corresponding standard deviation (%) | | | | | |
| 0.5 | 3 | 11.9 | 10.3 | 11.3 | 13.1 | 11.3 | 12.3 | 5 | 11 | 8 | 13 | 8 | 12 |
| 1 | 6 | 10.3 | 12.1 | 9.8 | 9.8 | 9.8 | 12.3 | 5 | 6 | 6 | 5 | 6 | 6 |
| 2 | 11 | 7.3 | 8.2 | 7.4 | 7.8 | 7.4 | 8.6 | 4 | 4 | 4 | 4 | 4 | 3 |
| 3 | 16 | 6 | 6.4 | 6.2 | 6.4 | 6.2 | 6.8 | 3 | 3 | 3 | 3 | 3 | 3 |
| 4 | 22 | 5.2 | 5.6 | 5.6 | 5.7 | 5.6 | 5.8 | 3 | 3 | 3 | 3 | 3 | 3 |
| 5 | 27 | 5 | 5.2 | 5.1 | 5.2 | 5.1 | 5.3 | 3 | 3 | 3 | 3 | 3 | 3 |
| % | Act | JHotdraw - Precision (%) | | | | | | Corresponding standard deviation (%) | | | | | |
| 0.5 | 3 | 10.8 | 16.9 | 10.1 | 12.6 | 10.1 | 19.9 | 2 | 8 | 8 | 6 | 8 | 8 |
| 1 | 6 | 9.7 | 10.8 | 9.3 | 11.2 | 9.3 | 13.4 | 2 | 5 | 3 | 4 | 3 | 6 |
| 2 | 11 | 7.1 | 7.9 | 7.1 | 7.7 | 7.1 | 8.7 | 2 | 3 | 2 | 2 | 2 | 3 |
| 3 | 17 | 5.6 | 5.9 | 6 | 6.3 | 6 | 6.8 | 2 | 2 | 2 | 2 | 2 | 3 |
| 4 | 23 | 4.9 | 5.3 | 5.3 | 5.6 | 5.3 | 5.9 | 1 | 2 | 2 | 2 | 2 | 2 |
| 5 | 28 | 4.8 | 4.9 | 4.9 | 5.1 | 4.9 | 5.4 | 2 | 2 | 2 | 2 | 2 | 2 |
| % | Act | QuickFIX/J - Precision (%) | | | | | | Corresponding standard deviation (%) | | | | | |
| 0.5 | 2 | 12.1 | 46.1 | 24.8 | 29.9 | 24.8 | 35 | 9 | 39 | 19 | 21 | 19 | 33 |
| 1 | 3 | 12.5 | 23.5 | 22.3 | 24.1 | 22.3 | 33.9 | 6 | 18 | 13 | 13 | 13 | 22 |
| 2 | 6 | 9.4 | 13.5 | 14.6 | 13.7 | 14.6 | 16.5 | 3 | 5 | 6 | 5 | 6 | 8 |
| 3 | 9 | 7.7 | 9.8 | 10.8 | 9.7 | 10.8 | 11.7 | 2 | 3 | 4 | 3 | 4 | 3 |
| 4 | 12 | 6.4 | 8.4 | 8.8 | 7.8 | 8.8 | 9.4 | 3 | 2 | 3 | 2 | 3 | 3 |
| 5 | 15 | 5.9 | 7.6 | 7.4 | 6.8 | 7.4 | 8.2 | 2 | 2 | 2 | 2 | 2 | 3 |
| % | Act | Overall Precision (%) | | | | | | Corresponding standard deviation (%) | | | | | |
| 0.5 | - | 11.7 | 23.1 | 15.1 | 18.1 | 15.1 | 21.3 | 6 | 28 | 14 | 17 | 14 | 22 |
| 1 | - | 10.8 | 15.2 | 13.5 | 14.6 | 13.5 | 19.2 | 5 | 12 | 10 | 10 | 10 | 16 |
| 2 | - | 7.9 | 9.7 | 9.5 | 9.6 | 9.5 | 11.0 | 3 | 5 | 5 | 5 | 5 | 6 |
| 3 | - | 6.4 | 7.3 | 7.6 | 7.4 | 7.6 | 8.3 | 3 | 3 | 4 | 3 | 4 | 4 |
| 4 | - | 5.5 | 6.4 | 6.5 | 6.3 | 6.5 | 6.9 | 3 | 3 | 3 | 3 | 3 | 3 |
| 5 | - | 5.2 | 5.9 | 5.7 | 5.7 | 5.7 | 6.2 | 3 | 3 | 3 | 3 | 3 | 3 |

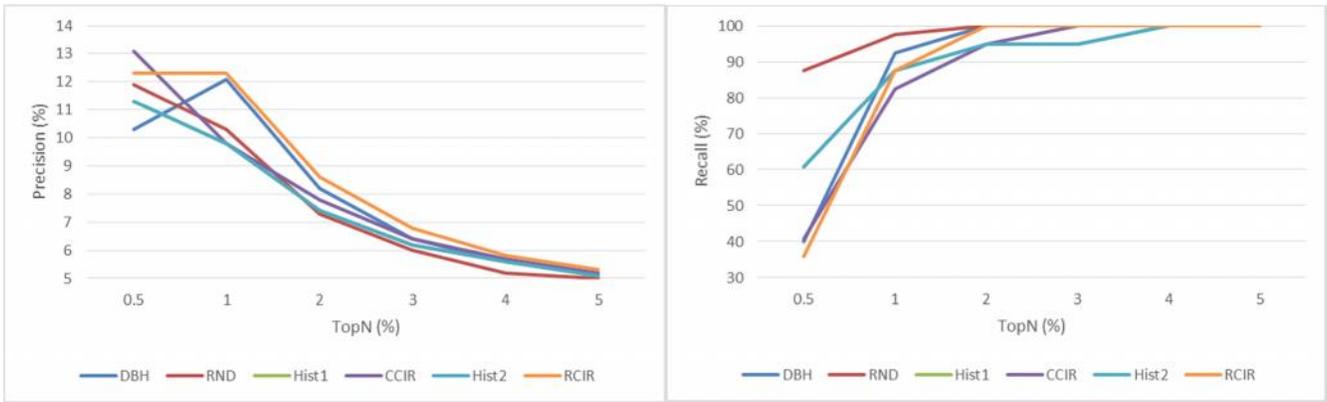

Figure 4. Average precision (left) and recall (right) of investigated heuristics for JEdit

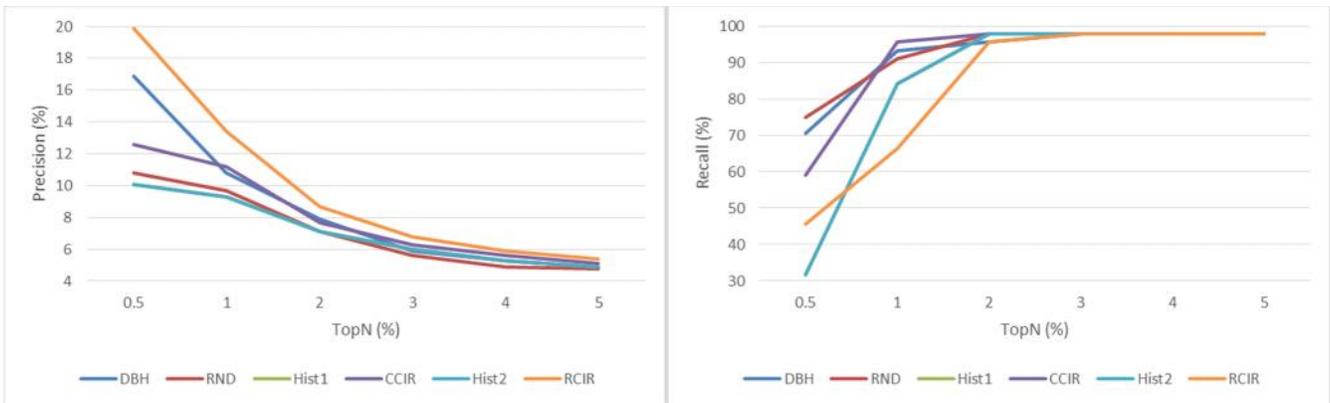

Figure 5. Average precision (left) and recall (right) of investigated heuristics for Jhotdraw

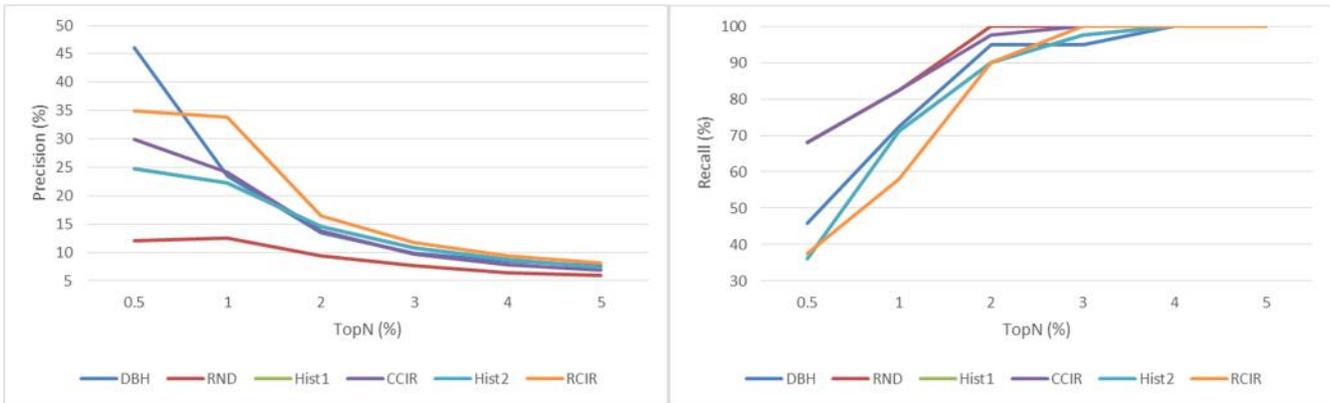

Figure 6. Average precision (left) and recall (right) of investigated heuristics for Quickfix/J

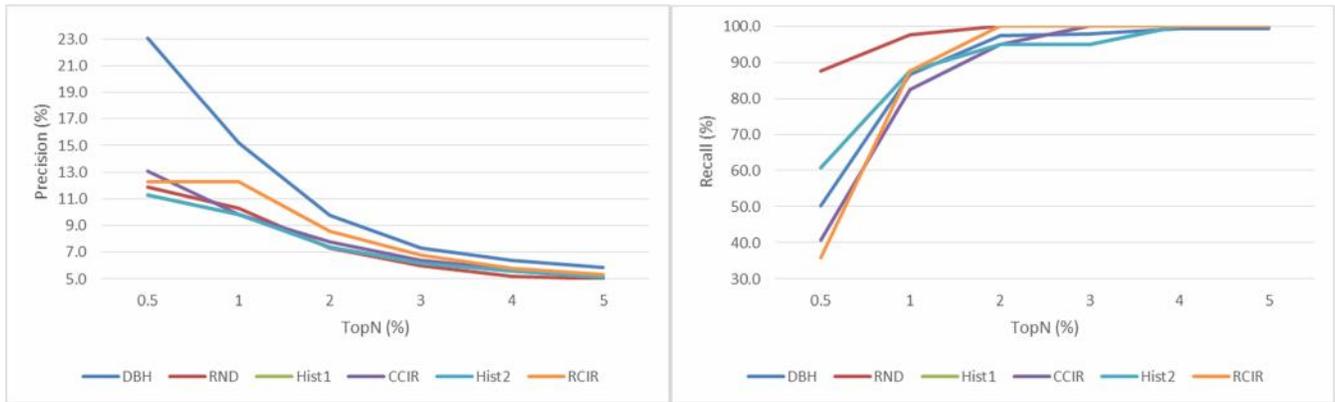

Figure 7. Average overall precision (left) and recall (right)

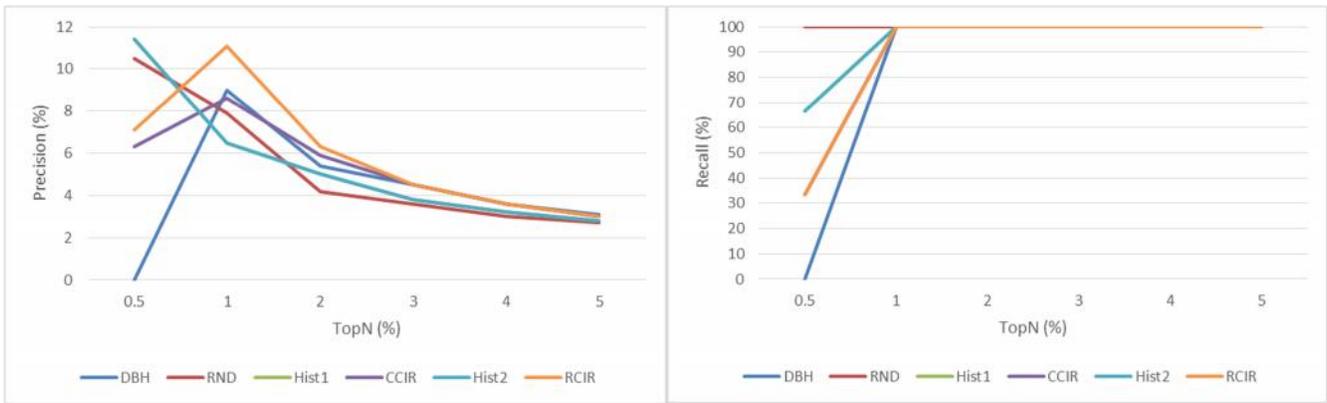

Figure 8. Median precision (left) and recall (right) of investigated heuristics for JEdit

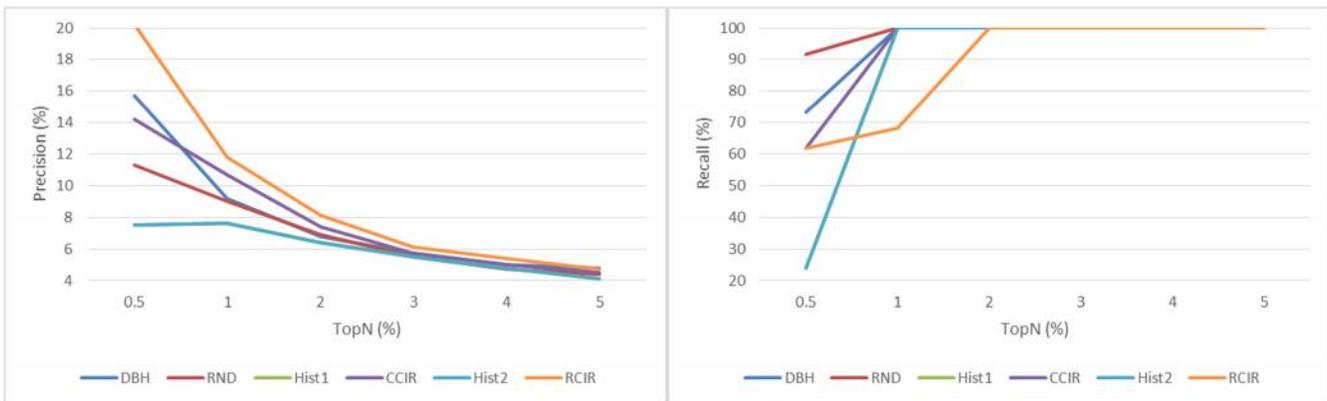

Figure 9. Median precision (left) and recall (right) of investigated heuristics for Jhotdraw

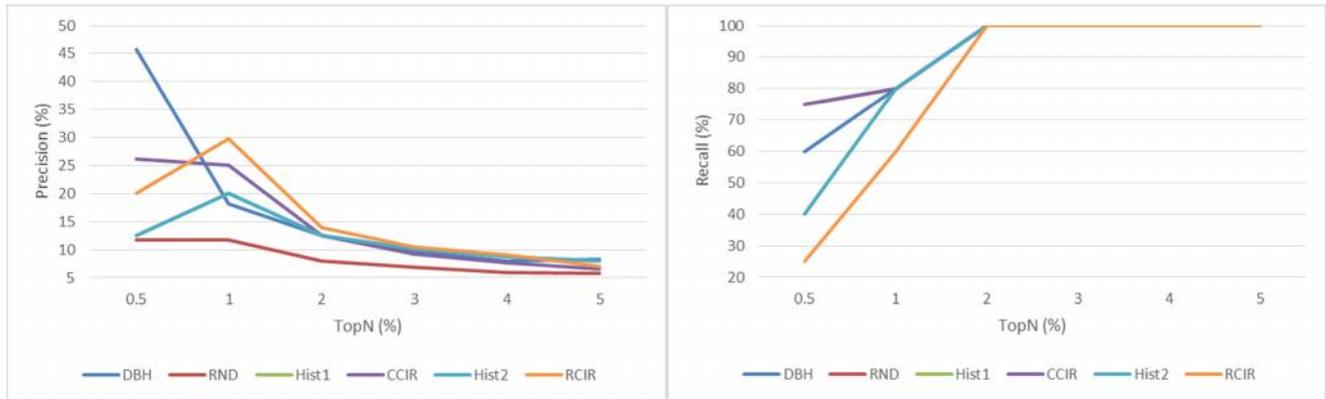

Figure 10. Median precision (left) and recall (right) of investigated heuristics for Quickfix/J

## B. Threats to Validity

Our study deals with the granularity of classes. Different results may be obtained for other granularities.

We evaluated only Java programs and different results may be obtained for other programming languages.

Some investigated heuristics such as Hist were adapted from AIA techniques. Those techniques are very different compared to IIA, as described in section III. In our solution, we convert them into propagation heuristics. Thus, some advantages of such heuristics may not be maintained during this conversion. It is possible that a different solution to adapt AIA techniques into IIA may provide different results.

Software repositories do not provide information about the initial impacted class. In the study, we considered every possible starting point of an impact analysis task. The performance of an IA using a particular propagation heuristic may vary slightly based on the selected IIC; however, this threat to validity is minor and does not affect the relative rankings of the investigated propagation heuristics.

Reenactment, as presented in this paper, may have certain built-in biases. Other empirical techniques, for example, empirical study of human developers, may provide different results.

## VI. CONCLUSIONS AND FUTURE WORK

In this study, we evaluated the performance of IIA combined with several propagation heuristics and a termination heuristic at the granularity of classes. To support the evaluation, we designed a novel empirical method based on the reenactment of IIA that simulates the actions of developers and reenact the past changes of open source projects, mined from software repositories. In general, IIA combined with the propagation heuristics that we explored, DBH, Hist1, CCIR, Hist2, and RCIR, performed better that other techniques known from the literature in terms of recall. However, all these heuristics fell short of expectations as they did not provide a convincing improvement when compared to the random inspection.

In view of this negative result, searching for good IIA heuristics is still on. In the future, we want to build on the experience from this research and search for more effective heuristics. For example, it is possible that classes ranked higher by multiple propagation heuristics are more likely to be impacted compared to the ones ranked higher only by a single propagation heuristic. We also plan to explore additional termination heuristics.

The methodology that we developed – reenactment – is giving us a clear comparison between different heuristics, and hence it will help us to assess whether the newly proposed heuristics are an improvement compared to the old ones. Hopefully in the future, a better IIA heuristics will emerge and help the developers to make more predictable and safe changes in software.

REFERENCES

[1] S. A. Bohner, "Software Change Impact Analysis," presented at the IEEE CS, 1996.

[2] V. Rajlich, Software Engineering: The Current Practice: CRC Press, 2011.

[3] S. S. Yau, J. S. Collofello, and T. M. MacGregor, "Ripple Effect Analysis of Software Maintenance," in Software engineering metrics I, S. Martin, Ed., ed: McGraw-Hill, Inc., 1993, pp. 71-82.

[4] A. E. Hassan and R. C. Holt, "Replaying development history to assess the effectiveness of change propagation tools," Empirical Software Engineering, vol. 11, pp. 335-367, 2006/09/01 2006.

[5] M. Petrenko and V. Rajlich, "Variable Granularity for Improving Precision of Impact Analysis," in IEEE 17th International Conference on Program Comprehension (ICPC '09), 2009, pp. 10-19.

[6] G. Tóth, P. Heged s, Á. Beszédes, T. Gyimóthy, and J. Jász, "Comparison of different impact analysis methods and programmer's opinion: an empirical study," presented at the Proceedings of the 8th

International Conference on the Principles and Practice of Programming in Java, Vienna, Austria, 2010.
[7]  M. Abi-Antoun, Y. Wang, E. Khalaj, A. Giang, and V. Rajlich, "Impact Analysis Based on a Global Hierarchical Object Graph," in 22nd IEEE International Conference on Software Analysis, Evolution, and Reengineering (SANER'15), Montréal, 2015, pp. 221-230.
[8]  H. Kagdi, M. Gethers, D. Poshyvanyk, and M. L. Collard, "Blending Conceptual and Evolutionary Couplings to Support Change Impact Analysis in Source Code," in 17th Working Conference on Reverse Engineering (WCRE'10), 2010, pp. 119-128.
[9]  M. Gethers, B. Dit, H. Kagdi, and D. Poshyvanyk, "Integrated Impact Analysis for Managing Software Changes," in 34th International Conference on Software Engineering (ICSE), 2012, pp. 430-440.
[10] H. Kagdi, M. Gethers, and D. Poshyvanyk, "Integrating Conceptual and Logical Couplings for Change Impact Analysis in Software," Empirical Software Engineering, vol. 18, pp. 933-969, 2013/10/01 2013.
[11] A. Aryani, "Predicting change propagation using domain-based coupling," Ph.D dissertation, Computer Science and Information Technology, RMIT University, 2013.
[12] X. Sun, B. Li, H. Leung, B. Li, and J. Zhu, "Static change impact analysis techniques: A comparative study," Journal of Systems and Software, vol. 109, pp. 137-149, 11// 2015.
[13] R. S. Arnold and S. A. Bohner, "Impact Analysis-Towards a Framework for Comparison," in Proceedings of Conference on Software Maintenance, Sept. 1993, pp. 292-301.
[14] S. Lehnert, "A taxonomy for software change impact analysis," presented at the Proceedings of the 12th International Workshop on Principles of Software Evolution and the 7th annual ERCIM Workshop on Software Evolution, Szeged, Hungary, 2011.
[15] B. Li, X. Sun, H. Leung, and S. Zhang, "A survey of code-based change impact analysis techniques," Software Testing, Verification and Reliability, vol. 23, pp. 613-646, 2013.
[16] H. Malik and A. E. Hassan, "Supporting Software Evolution Using Adaptive Change Propagation Heuristics," in 24th IEEE International Conference on Software Maintenance (ICSM 2008), 2008, pp. 177-186.
[17] M. B. Zanjani, G. Swartzendruber, and H. Kagdi, "Impact Analysis of Change Requests on Source Code Based on Interaction and Commit Histories," presented at the Proceedings of the 11th Working Conference on Mining Software Repositories, Hyderabad, India, 2014.
[18] B. Dit, M. Wagner, S. Wen, W. Wang, M. L. Vásquez, D. Poshyvanyk, et al., "ImpactMiner: A Tool for Change Impact Analysis," in ICSE Companion, 2014, pp. 540-543.
[19] M. Borg, K. Wnuk, B. Regnell, and P. Runeson, "Supporting Change Impact Analysis Using a Recommendation System: An Industrial Case Study in a Safety-Critical Context," IEEE Transactions on Software Engineering, vol. 43, pp. 675-700, 2017.
[20] V. Musco, A. Carette, M. Monperrus, and P. Preux, "A Learning Algorithm for Change Impact Prediction: Experimentation on 7 Java Applications," arXiv preprint arXiv:1512.07435, 2015.
[21] V. Musco, A. Carette, M. Monperrus, P. Preux, and F. Lille, "A Learning Algorithm for Change Impact Prediction," in 5th International Workshop on Realizing Artificial Intelligence Synergies in Software Engineering, 2016.
[22] E. Khalaj and M. Abi-Antoun, "Inferring Ownership Domains from Refinements," presented at the Proceedings of the 17th ACM SIGPLAN International Conference on Generative Programming: Concepts and Experiences, Boston, MA, USA, 2018.
[23] M. Acharya and B. Robinson, "Practical Change Impact Analysis Based on Static Program Slicing for Industrial Software Systems," presented at the Proceedings of the 33rd International Conference on Software Engineering, Waikiki, Honolulu, HI, USA, 2011.
[24] H. Cai, S. Jiang, R. Santelices, Y.-J. Zhang, and Y. Zhang, "SENSA: Sensitivity Analysis for Quantitative Change-Impact Prediction," in IEEE 14th International Working Conference on Source Code Analysis and Manipulation (SCAM), 2014, pp. 165-174.
[25] A. Beszedes, T. Gergely, J. Jasz, G. Toth, T. Gyimothy, and V. Rajlich, "Computation of Static Execute After Relation with Applications to Software Maintenance," in IEEE International Conference on Software Maintenance (ICSM 2007), 2007, pp. 295-304.
[26] H. Cai and R. Santelices, "Abstracting Program Dependencies Using the Method Dependence Graph," in 2015 IEEE International Conference on Software Quality, Reliability and Security (QRS), 2015, pp. 49-58.
[27] C. Dorman and V. Rajlich, "Software Change in the Solo Iterative Process: An Experience Report," in 2012 Agile Conference, 2012, pp. 21-30.
[28] J. Buckner, J. Buchta, M. Petrenko, and V. Václav, "JRipples: A Tool for Program Comprehension during Incremental Change," presented at the Proceedings of the 13th International Workshop on Program Comprehension, 2005.
[29] B. Li, X. Sun, and H. Leung, "Combining concept lattice with call graph for impact analysis," Advances in Engineering Software, vol. 53, pp. 1-13, 11 2012.
[30] M. Petrenko, "On use of dependency and semantics information in incremental change," Ph.D dissertation, Wayne State University, 2009.
[31] L. C. Briand, J. Wüst, and H. Lounis, "Using Coupling Measurement for Impact Analysis in Object-Oriented Systems," in IEEE International Conference on Software Maintenance (ICSM '99), 1999, pp. 475-482.
[32] D. Poshyvanyk, A. Marcus, R. Ferenc, and T. Gyimóthy, "Using information retrieval based coupling measures for impact analysis," Empirical Software Engineering, vol. 14, pp. 5-32, 2009.
[33] G. Antoniol, G. Canfora, G. Casazza, and A. De Lucia, "Identifying the Starting Impact Set of a Maintenance Request: a Case Study," in European Conference on Software Maintenance and Reengineering, 2000, pp. 227-230.
[34] H. Kagdi, M. L. Collard, and J. I. Maletic, "A Survey and Taxonomy of Approaches for Mining Software Repositories in the Context of Software Evolution," Journal of Software Maintenance and Evolution: Research and Practice, vol. 19, pp. 77-131, 2007.
[35] T. Zimmermann, A. Zeller, P. Weissgerber, and S. Diehl, "Mining Version Histories to Guide Software Changes," IEEE Transactions on Software Engineering, vol. 31, pp. 429-445, 2005.
[36] F. K. Hwang and D. S. Richards, "Steiner Tree Problems," Networks, vol. 22, pp. 55-89, 1992.
[37] H. Takahashi and A. Matsuyama, "An approximate solution for the Steiner problem in graphs," Mathematica Japonica, vol. 24, pp. 573-577, 1980.